\documentclass[times,twocolumn,final,authoryear]{elsarticle}

\usepackage{jasr}
\usepackage{framed,multirow}

\usepackage{amssymb}
\usepackage{latexsym}

\usepackage[switch]{lineno}

\usepackage{url}
\usepackage{xcolor}
\definecolor{newcolor}{rgb}{.8,.349,.1}

\usepackage[citebordercolor=white]{hyperref}

\journal{Advances in Space Research}

\begin{document}

\verso{Pohjolainen, Talebpour Sheshvan, and Monstein}

\begin{frontmatter}

\title{Separating the effects of earthside and far side solar events. A case study}

\author[1]{Silja \snm{Pohjolainen}\corref{cor1}}
\cortext[cor1]{Corresponding author: silpoh@utu.fi}
\author[2]{Nasrin \snm{Talebpour Sheshvan}}
\author[3]{Christian \snm{Monstein}}

\address[1]{Tuorla Observatory, Department of Physics and Astronomy,
  University of Turku, 20014 Turun yliopisto, Finland}
\address[2]{Department of Physics and Astronomy, University of Turku,
            20014 Turun yliopisto, Finland}
\address[3]{Istituto ricerche solari Aldo e Cele Dacc{\`o} (IRSOL),
  Faculty of Informatics, Universit{\`a} della Svizzera italiana (USI),
           CH-6605 Locarno, Switzerland}

\received{}
\finalform{}
\accepted{}
\availableonline{}
\communicated{}

\begin{abstract}
On 8 November 2013 a halo-type coronal mass ejection (CME) was observed,
together with flares and type II radio bursts, but the association
between the flares, radio bursts, and the CME was not clear.
Our aim is to identify the origin of the CME and its direction of
propagation, and to exclude features that were not connected to it.
On the Earth-facing side, a GOES C5.7 class flare occurred close to the
estimated CME launch time, followed by an X1.1 class flare.
The latter flare was associated with an EUV wave and metric type II
bursts. On the far side of the Sun, a filament eruption, EUV dimmings, and
ejected CME loops were observed by
imaging instruments onboard the Solar TErrestrial RElations Observatory
(STEREO) spacecraft that were viewing the backside of the Sun. The STEREO
radio instruments observed an interplanetary (IP) type II radio burst at
decameter-hectometric wavelengths, which was not observed by the radio
instrument onboard the Wind spacecraft located at L1 near Earth.
We show that the halo CME originated from the eruption on the far side
of the Sun, and that the IP type II burst was created by a shock wave
ahead of the halo CME. The radio burst remained unobserved from the
earthside, even at heliocentric source heights larger
than 9 solar radii. During the CME propagation, the X-class flare eruption
caused a small plasmoid ejection earthward, the material of which was
superposed on the earlier CME structures observed in projection.
The estimated heights of the metric type II burst match well with the
EUV wave launched by the X-class flare. As this radio emission did not
continue to lower frequencies, we conclude that the shock wave did not
propagate any further. Either the shock driver died out, as a blast
wave, or the driver speed no longer exceeded the local Alfven speed.
\end{abstract}


\end{frontmatter}


\section{Introduction}
\label{intro}

Solar flares and eruptions are sometimes associated with propagating shock
waves and coronal mass ejections (CMEs). The flare intensities, durations,
eruption energetics, and mass sizes vary, and they do not necessarily depend
on each other \citep{kawabata2018,kahler2022}.
Many events are observed in projection, which makes it difficult
to estimate source heights \citep{kwon2015,balmaceda2018,vourlidas2020}.
Also separating far side events from Earth view events can be difficult
for halo CMEs, as we cannot conclude the propagation direction if we only
see  a symmetrical halo surrounding the solar disc. 

Flare-accelerated particles create emissions that can be observed especially
in hard X-rays and microwaves. Microwave emission is typically due to
trapped electrons gyrating in the closed magnetic field (gyrosynchrotron
emission), and hard X-rays are created when some of these electron
populations collide with dense material (often at the footpoints of
flare loops). Plasma emission, on the other hand, is created when a transient
propagates through a medium and causes oscillations that turn into radio
waves at the local plasma frequency. Reviews on the various aspects of solar
radio emission associated with solar events have been published by,
e.g.,  \citet{bastian98}, \citet{nindos2008}, and \citet{pick2008}.

Observations at plasma frequencies include type III bursts which are created
by electron beams that follow open magnetic field lines out from the Sun
and into the interplanetary (IP) space. Propagating shock waves can create
type II radio bursts, but the shock origin cannot always be determined
\citep{fulara2021,chernov2021}. Flare (blast) waves, driven waves, and bow
shocks are all possible in connection with solar eruptions
\citep{warmuth2007}. Shock drivers can be fast CMEs or plasmoids,
or fast expansion at the CME flanks. For bow shocks, the transient speed
must exceed the local Alfven speed.

Radio emission at plasma frequency is useful in a sense that it can be
used to determine source heights. As plasma frequency depends only on the
local electron density, we can use coronal density models to obtain
estimates of distance from the solar surface. The conversion of radio
frequency to atmospheric height has been reviewed in, e.g.,
\cite{pohjolainen2007}. 

The research questions addressed in the present analysis are:
Can we distinguish which parts of a halo CME are caused by far side eruption
and which parts belong to frontside eruption? Can we exclude one or the other,
based on observed flares and shock waves? Can radio emission help in separating
source origins? We present a case study of multiple emissions and features that
originated from active regions located on the earthside and the far side
of the Sun, with possible associations to a halo CME. 


\section{Data analysis}
\label{data}

We have analysed solar eruptions that occurred on 8 November 2013.
Eruptions were observed on the earthside and the far side of the Sun,
see Table \ref{table1}. A halo CME was observed during the
investigated time period of 00:00--06:00 UT. 

For the analysis we used solar EUV images provided by the Atmospheric Imaging
Assembly (AIA) \citep{lemen2012} onboard the Solar Dynamics Observatory (SDO), and
the Extreme Ultraviolet Imager (EUVI) onboard the STEREO A and B spacecrafts
\citep{wuelser2004}. Coronagraph images and associated data products were
obtained from the CDAW LASCO CME Catalog at \url{https://cdaw.gsfc.nasa.gov}
that includes SOHO LASCO and STEREO A and B coronagraph images.
Full-disc soft X-ray flux were provided by GOES satellites.  

Radio data at decimeter-metric wavelengths include dynamic spectra from
several e-CALLISTO Network stations \citep{benz2005},  
\url{https://www.e-callisto.org/}, and from the Radio Solar Telescope
Network (RSTN) Learmonth station,
\url{https://www.ngdc.noaa.gov/stp/space-weather/solar-data/solar-features/solar-radio/rstn-spectral}. 
For interplanetary radio emission at decameter-hectometric waves we used
dynamic spectra from WAVES onboard the Wind satellite \citep{bougeret95},
and SWAVES spectra from the STEREO A and B satellites \citep{bougeret2008}.
These can be accessed in the CDAW LASCO CME Catalog. 
Microwave single frequency data were obtained from the Nobeyama radio
polarimeters (NoRP) at \url{http://solar.nro.nao.ac.jp/norp/}.


\begin{table}
\centering
\caption{Flare/eruption times, locations, and intensities on 8 November 2013
  for the time period of 00:00--06:00 UT. Two flares were listed to occur
  on the visible solar disc and a filament eruption was observed on the far
  side of the Sun. End times have been estimated from the GOES flux curves
  (not as in the NOAA flare listings). Only one CME was listed for this time
  period. }
\begin{tabular}{llll}
\hline
Start    & End       & Location & GOES class/  \\
time     & time      &          & or type \\
\hline
02:25 UT &           & S20W180  & Filament eruption\\
         &           &          & southward \\
02:33 UT & 03:10 UT  & S18W25   & C5.7  \\
04:20 UT & 04:45 UT  & S14E15   & X1.1 \\
\hline
\end{tabular}
\label{table1}
\end{table}


\begin{figure}
  \centering
\includegraphics[width=0.5\textwidth]{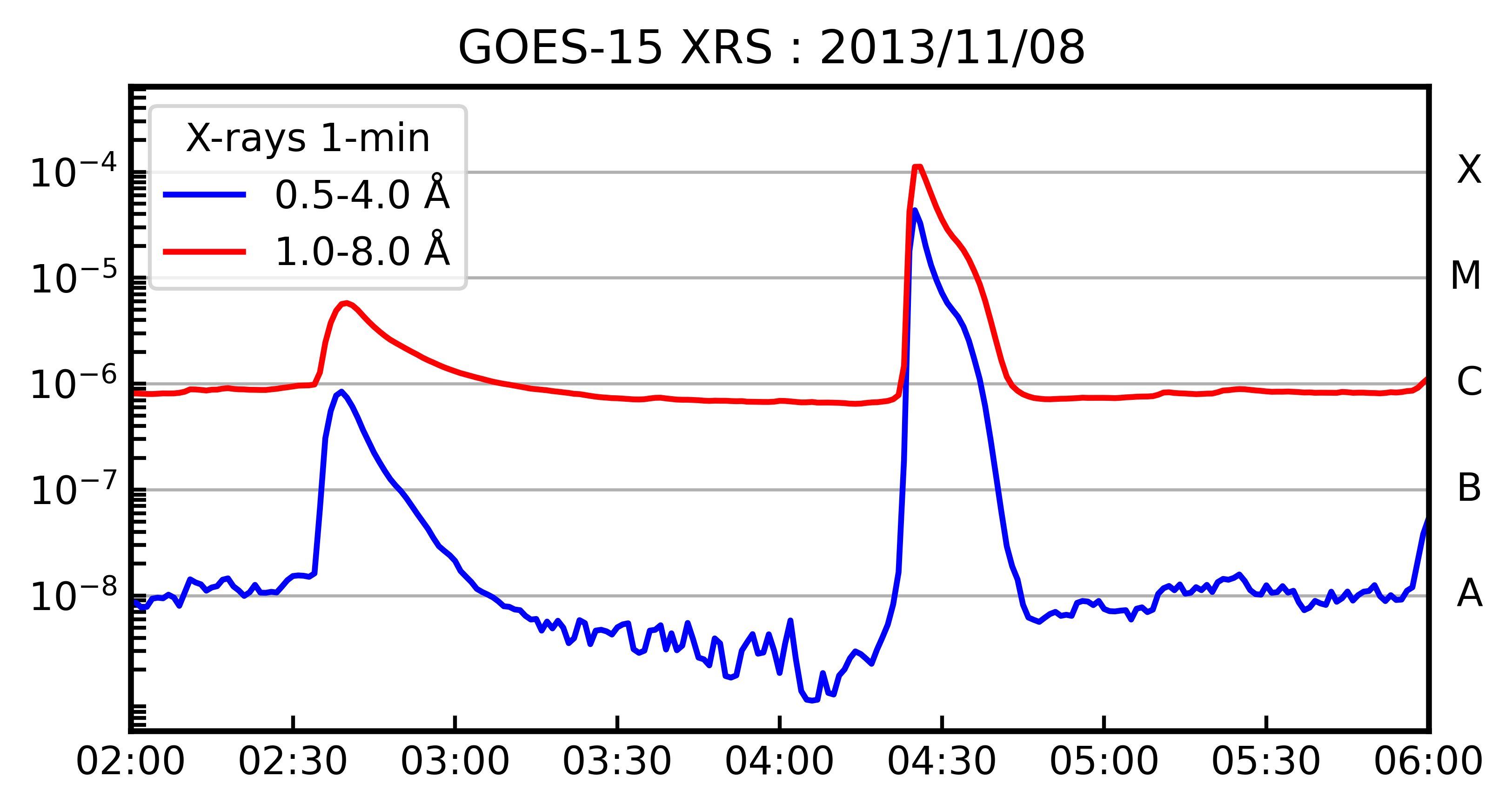}  
\caption{GOES soft X-ray flux curve shows the C5.7 and X1.1 class flares.
   }
\label{fig:GOES}
\end{figure}

\begin{figure}
  \centering
  \includegraphics[width=0.4\textwidth]{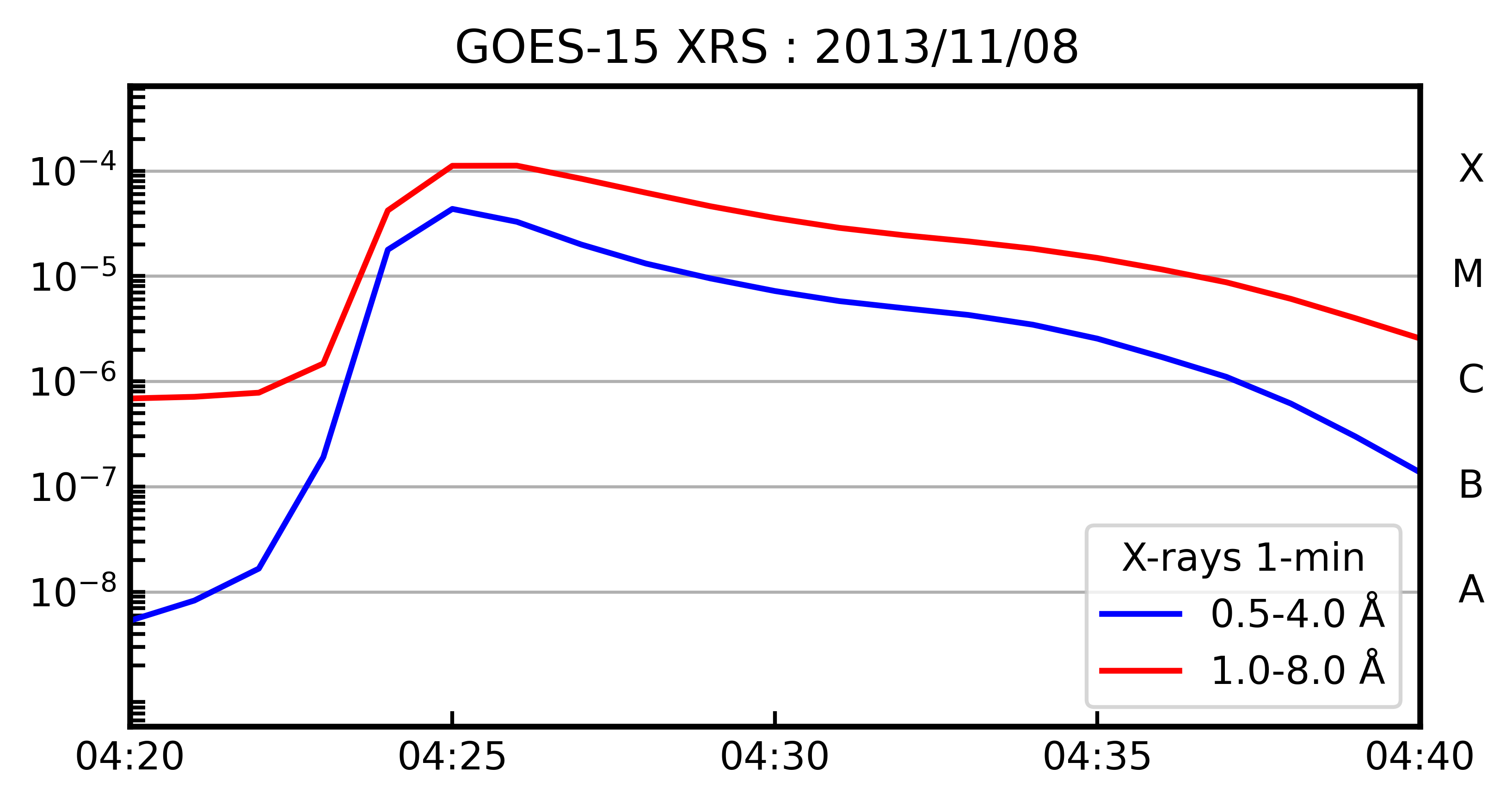}
  \includegraphics[width=0.4\textwidth]{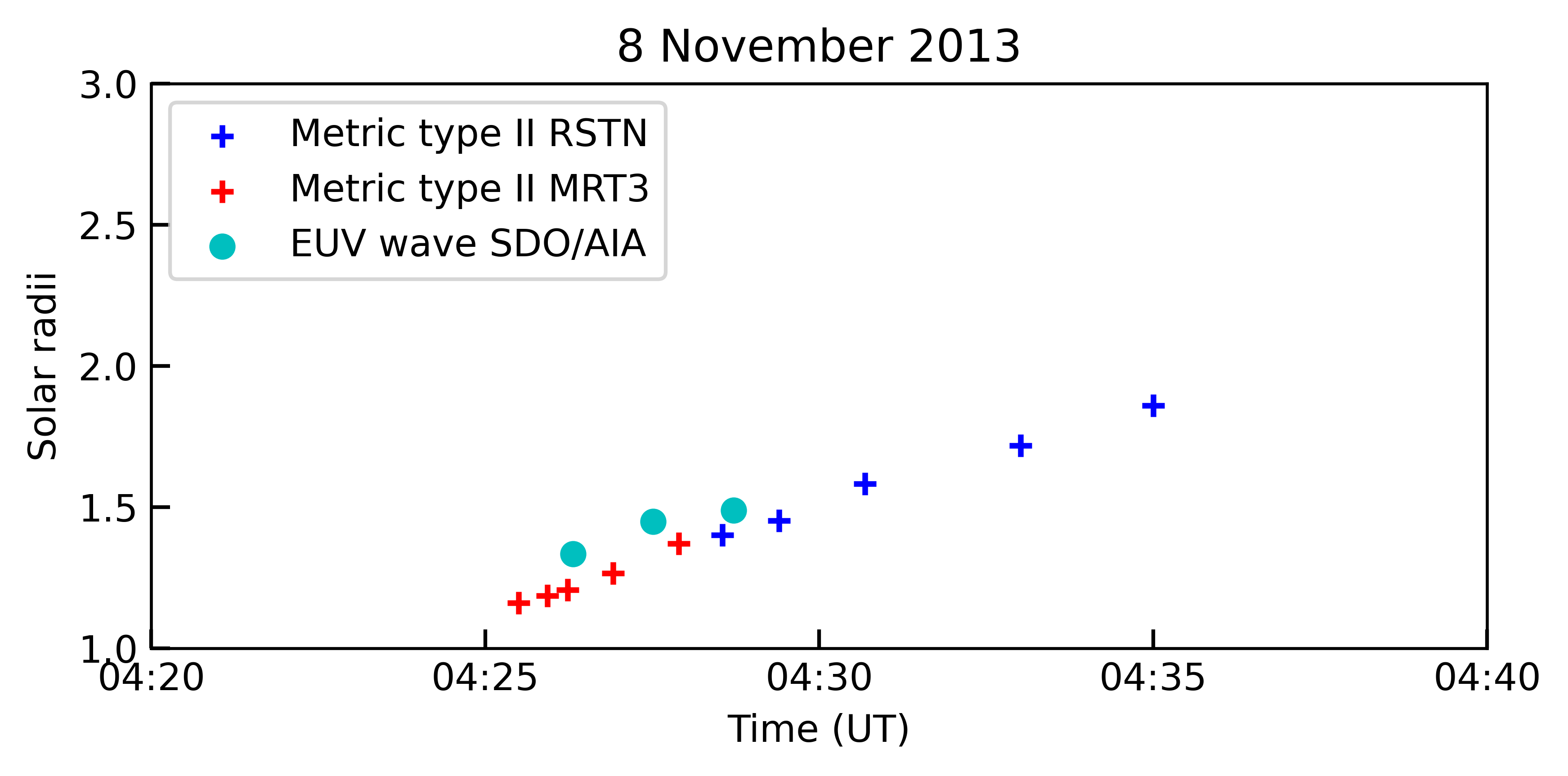}
\includegraphics[width=0.4\textwidth]{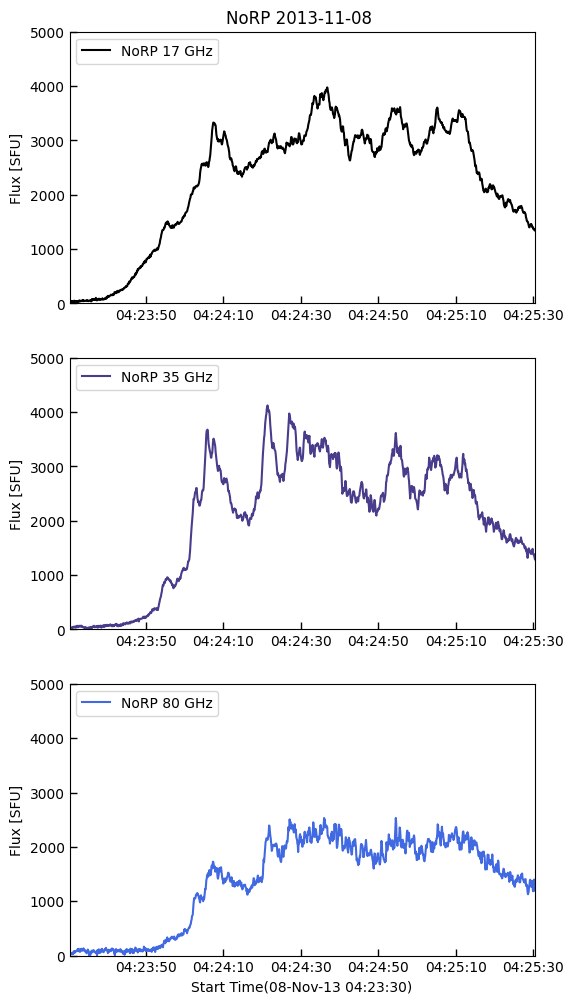}
\caption{Top: GOES X-rays from the X1.1 class flare.
  Middle: Estimated heliocentric distances/heights for the EUV wave
  (circles), and the metric type II burst (crosses) observed
  by e-CALLISTO MRT3 at 450--50 MHz and RSTN Learmonth at 180--25 MHz.
  The calculated radio source heights are listed in Table \ref{table2}.
  Bottom three plots: Radio flux densities from the Nobeyama polarimeters
  at 17, 35, and 80 GHz, in the flare impulsive phase.
}
\label{fig:heights-low}
\end{figure}

\begin{figure}
  \centering
  \includegraphics[width=0.5\textwidth]{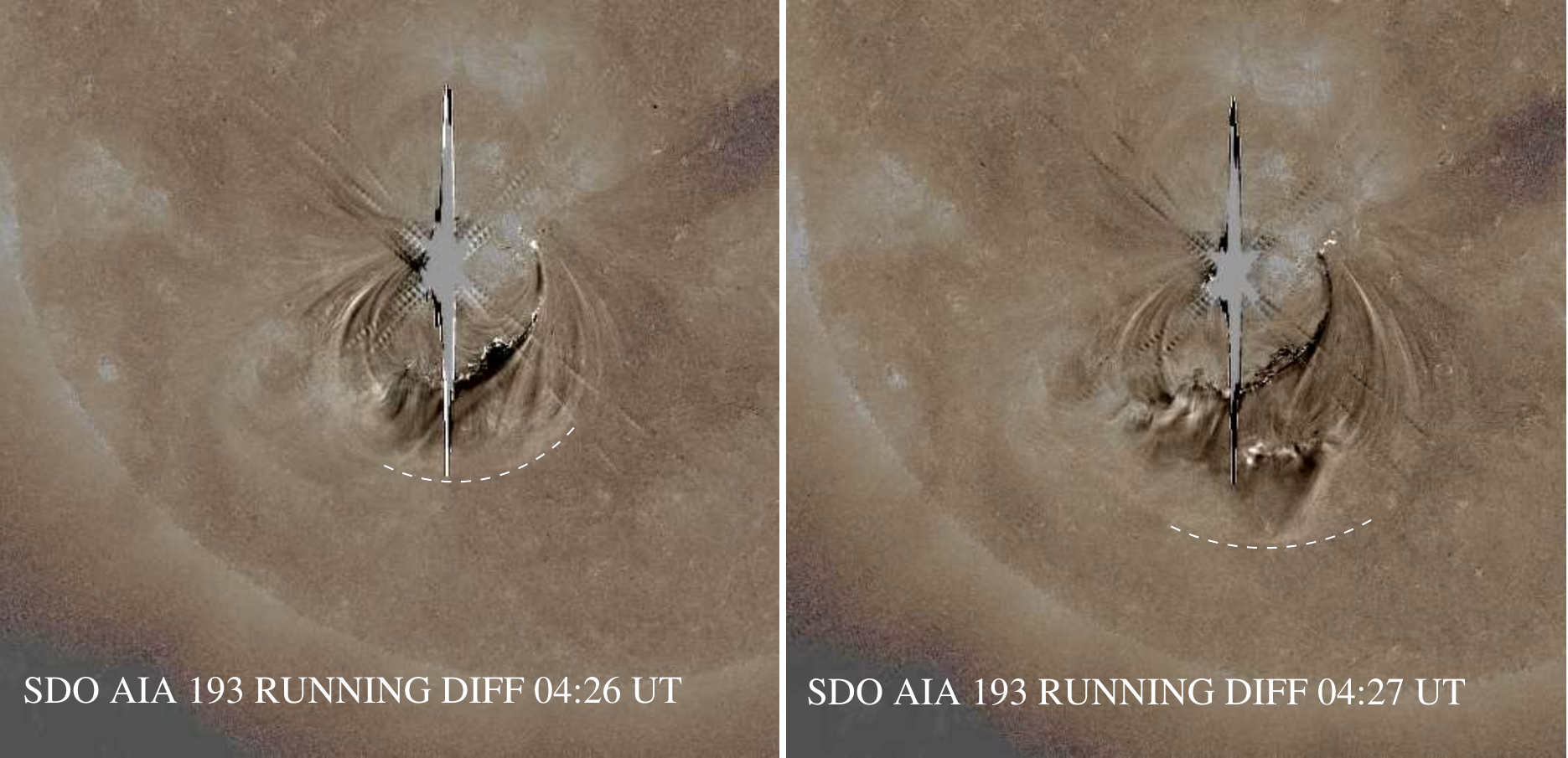}
  \caption{SDO/AIA 193 \AA \, difference images that show the evolution of
  the EUV wave that followed the GOES X1.1 class flare located at S14E15.     
  }
  \label{fig:aia-runningdiff}
\end{figure}


\subsection{Flares and eruptions}
\label{flares}

The first flare observed by GOES on 8 November 2013 is GOES class C5.7,
start time at 02:33 UT, and duration of approximately 37 minutes, see 
Fig. \ref{fig:GOES}. The flare is reported to have occurred
at location S18W25 in NOAA active region (AR) 11891, and a corresponding
small-scale loop brightening can be observed in the SDO/AIA images.

It is followed by a GOES X1.1 class flare in another region, in AR 11890
at location S14E15. The flare is reported to have a start time at 04:20 UT,
flux peak time at 04:26 UT, and we estimated that the flare duration was
approximately 25 minutes (Fig. \ref{fig:GOES}). 

This X-class flare was well-observed at microwaves and mm-waves, the peak
flux reaching 4000 sfu at 17--35 GHz, and 2200 sfu at 80 GHz
(Fig. \ref{fig:heights-low}). The flux began to rise at 04:23:40~UT, and
multiple flux peaks were visible until 04:25:12 UT. Unfortunately no hard
X-ray observations are available for this flare, but the high flux density at
80 GHz indicates particle acceleration to high energies. 

Starting at 04:24 UT, decimeter-metric radio emission was observed by
several ground-based stations. These emission structures are described
in more detail in Section \ref{radio-emission}

\begin{figure}
\centering
\includegraphics[width=0.5\textwidth]{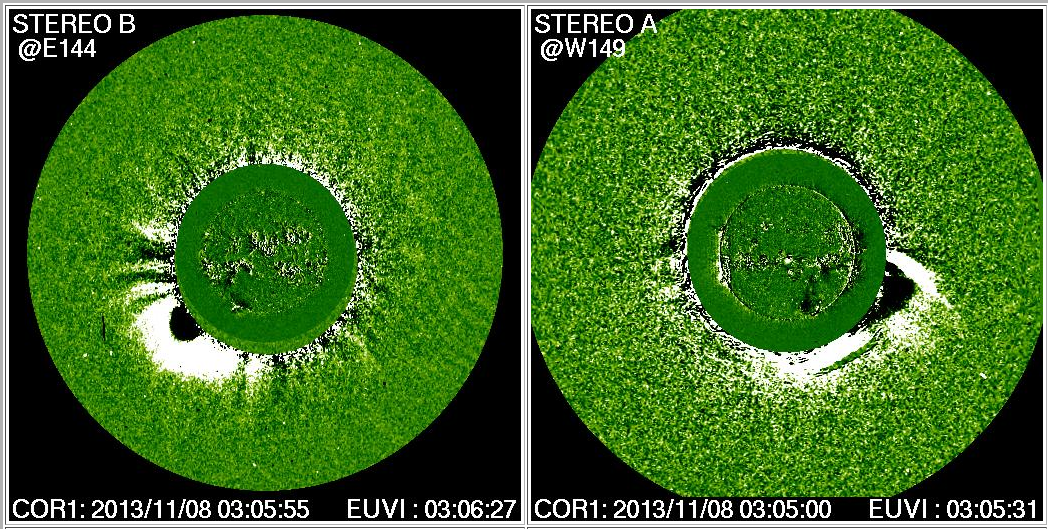}\\
\vspace{10mm}
\includegraphics[width=0.4\textwidth]{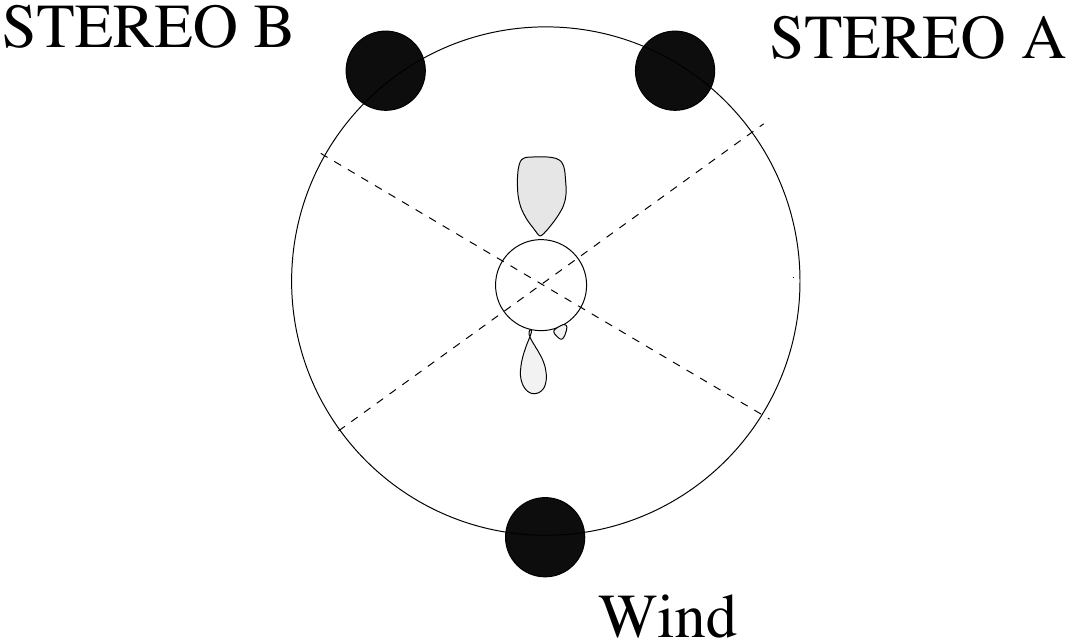}
\caption{Dimming regions and the launch of a CME on the far side of the Sun 
  (STEREO A and B observations). The X-class flare on the earthside
  originated from an active region at E15, the C-class flare was a brightening
  at W25, and the far side eruption originated from a region near W180.
  Cartoon shows the locations of these eruptions, together with STEREO A and B
  positions. SOHO/LASCO and SDO have the same observing direction as Wind.
  The earthside flaring active regions at low heights were unobservable
  from STEREO A and B EUV imagers.
  }
\label{fig:backside-euvi}
\end{figure}

The X-class flare was associated with a bubble-like EUV wave. The direction
of the wave was southward from the AR, see Fig. \ref{fig:aia-runningdiff}.
We measured the wave distances (dashed lines shown in
Fig. \ref{fig:aia-runningdiff}) from the AR center at S14E15, assuming
that the distance on the disc compares to the wave height.
The distance increased from 0.333 R$_{\odot}$ at 04:26 UT to 0.489 R$_{\odot}$
at 04:28 UT, indicating a speed of approximately 900~km~s$^{-1}$.
In Fig. \ref{fig:heights-low} the distances are shown in heliocentric
heights, for comparison purposes.

On the far side of the Sun a filament eruption was observed
at 02:25 UT. The most visible part of the filament was already at some
distance from the AR location at S20W180, and it was directed toward south.
STEREO A and B EUVI images show several dimming regions in the same direction
(Fig. \ref{fig:backside-euvi}).
Later on, dimmings appeared also in the location of the AR loops. 

\begin{figure*}
\centering
\includegraphics[width=0.95\textwidth]{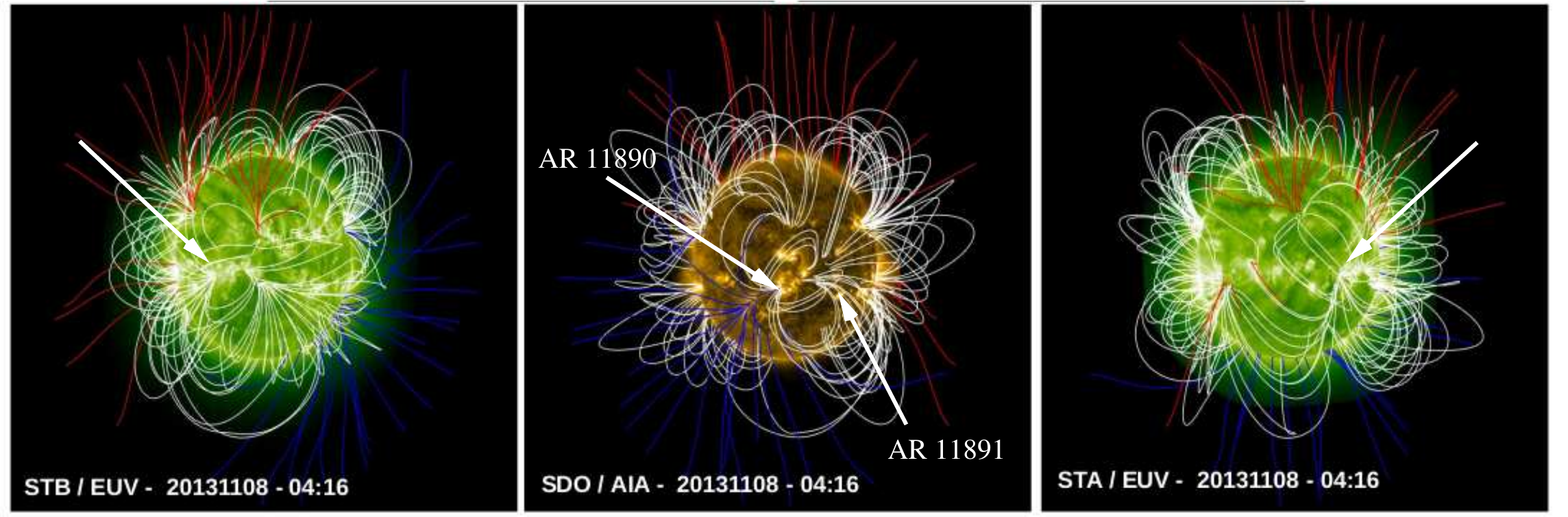}
\caption{Magnetic field lines calculated with the potential field
  source surface (pfss) model, shown in STEREO B, SDO, and STEREO A view.
    Active regions 11890 and 11891 are marked in the SDO view map, and
    white arrows mark the erupting region on the far side, near S20W180,
    in STEREO A and B views.
  Closed field lines are shown in white and open field lines in red and blue
  (opposite polarities).
  }  
\label{fig:pfss}
\end{figure*}

Potential field source surface (pfss) magnetic field lines were calculated
using synoptic maps, as on the far side we have no direct magnetic
field measurements. The field lines are shown both for Earth view and
far side views in Fig. \ref{fig:pfss}.
They show that open field lines exist on the south-east side of AR 11890,
but all other eruption regions lack open field lines in their vicinity.
Flare-accelerated electrons would need open field lines to stream out
from the Sun. Lack of type III bursts, especially in the interplanetary
space, would also suggest that particles did not have access to any.
For connectivity to observing particle detectors, it is useful to know
where the open field lines are directed.
  
We note that the flaring regions AR 11890 and AR 11891 on the disc
could not have been observed from STEREO A and B spacecraft,
as they were located behind the limb in their field of views,
see cartoon in Fig. \ref{fig:backside-euvi}.

\subsection{Radio emission}
\label{radio-emission}

Decimeter-metric radio emission started as narrow-width type III burst
emission at $\approx$\,600 MHz at 04:24 UT. At 04:25~UT a clear type
II burst emission lane became visible at 160 MHz, and it could be followed
down to 30 MHz (e-CALLISTO and RSTN Learmonth dynamic spectra), see
Fig. \ref{fig:lear-times}. 
The burst showed fundamental and harmonic emission lanes that were also
band-split. The real start time of the burst is unclear, due to the
overlap with other emission structures.

The metric type II burst did not continue to IP space (note that there
is a data gap between 25 and 15 MHz, and therefore the end time and
frequency cannot be determined). The disappearance of the type II burst
may indicate that the shock wave did not propagate any further. 
For example, a blast wave shock would eventually loose it's energy
and stop. For a bow shock a rise in the local Alfven speed would make
the shock die out if the transient speed no longer exceeded the local
magnetosonic speed (note that the transient could still continue to
move out at the same speed). To identify shock origin or shock driver
is difficult without knowing the local plasma conditions, and without
radio imaging at these wavelengths.

We estimated the metric type II burst heights from the \mbox{e-CALLISTO} MRT3
(Mauritius Radio Telescope, antenna number 3) and RSTN Learmonth dynamic
spectra using the 'hybrid' coronal density model of \cite{vrsnak04}.
The derived instantaneous type II shock speeds
were in the range of 670--980 km s$^{-1}$. As we do not know the local
coronal densities but rely on the density model, this can be taken as
an approximation of the shock speed. However, the radio emission heights
and speeds are not far away from those of the EUV wave observed during
the X-class flare. The metric type II burst heights are shown in
Fig. \ref{fig:heights-low} (cross symbols in the bottom plot),
with the EUV wave distance/heights (circles). Type II burst
frequencies and corresponding heights are listed in Table \ref{table2}.  


\begin{figure}
  \includegraphics[width=0.45\textwidth]{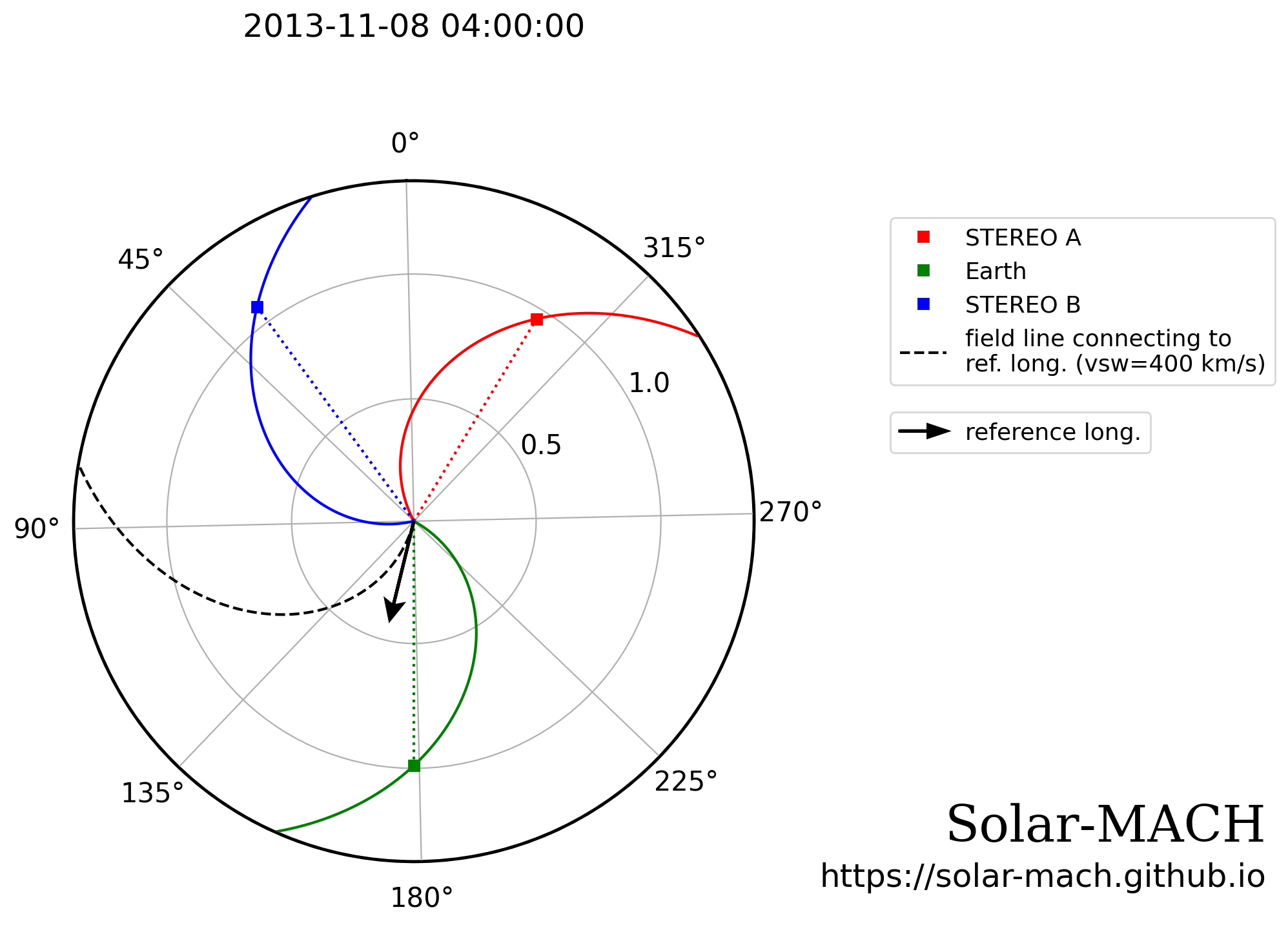}
  \includegraphics[width=0.3\textwidth]{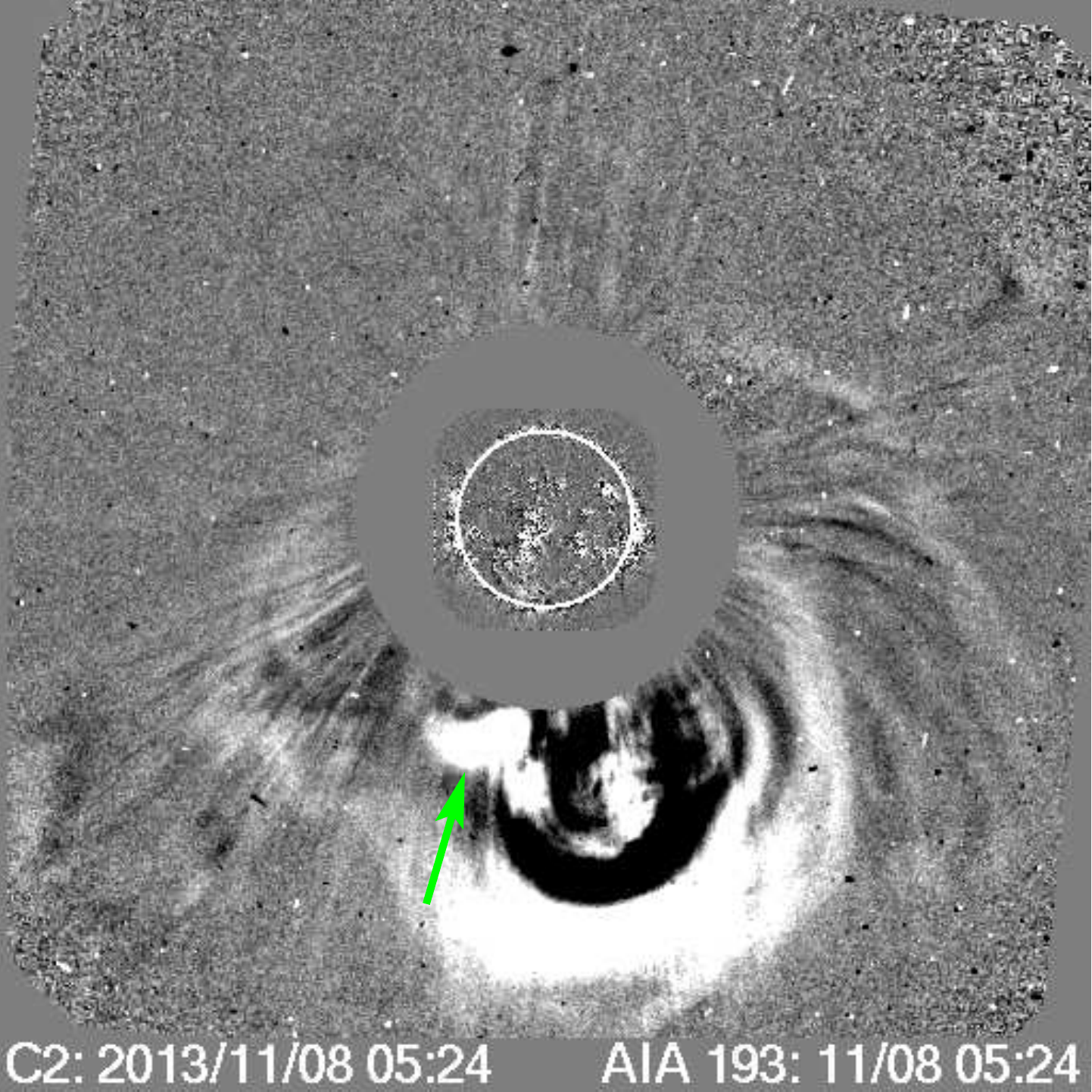}
  \includegraphics[width=0.3\textwidth]{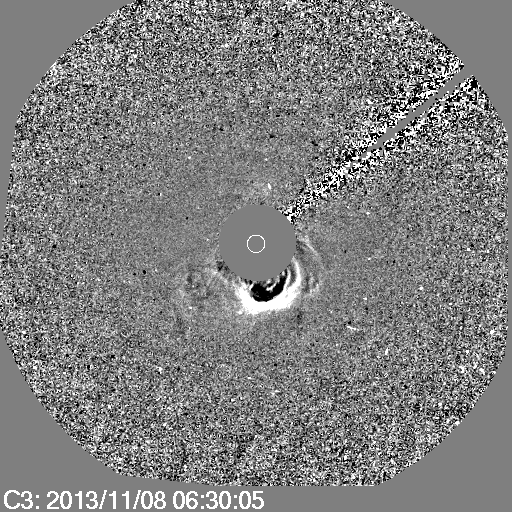}
  \caption{Solar-MACH plot (prepared with the Solar MAgnetic Connection HAUS
    tool) shows the field line connecting to the reference longitude of the
    X1.1 class flare located at S14E15 (indicated with the dashed line, top). 
    SOHO/LASCO C2 observation of the CME at 05:24 UT (middle).
    A separate plasmoid is visible, indicated with the green arrow.
    SOHO/LASCO C3 observation of the CME at 06:30 UT (bottom). During
    06:00--07:00 UT an IP type II burst was observed by STEREO A and B, but
    not by Wind. 
    }
  \label{fig:cme-flare}
\end{figure}

\begin{figure}
  \centering
  \includegraphics[width=0.5\textwidth]{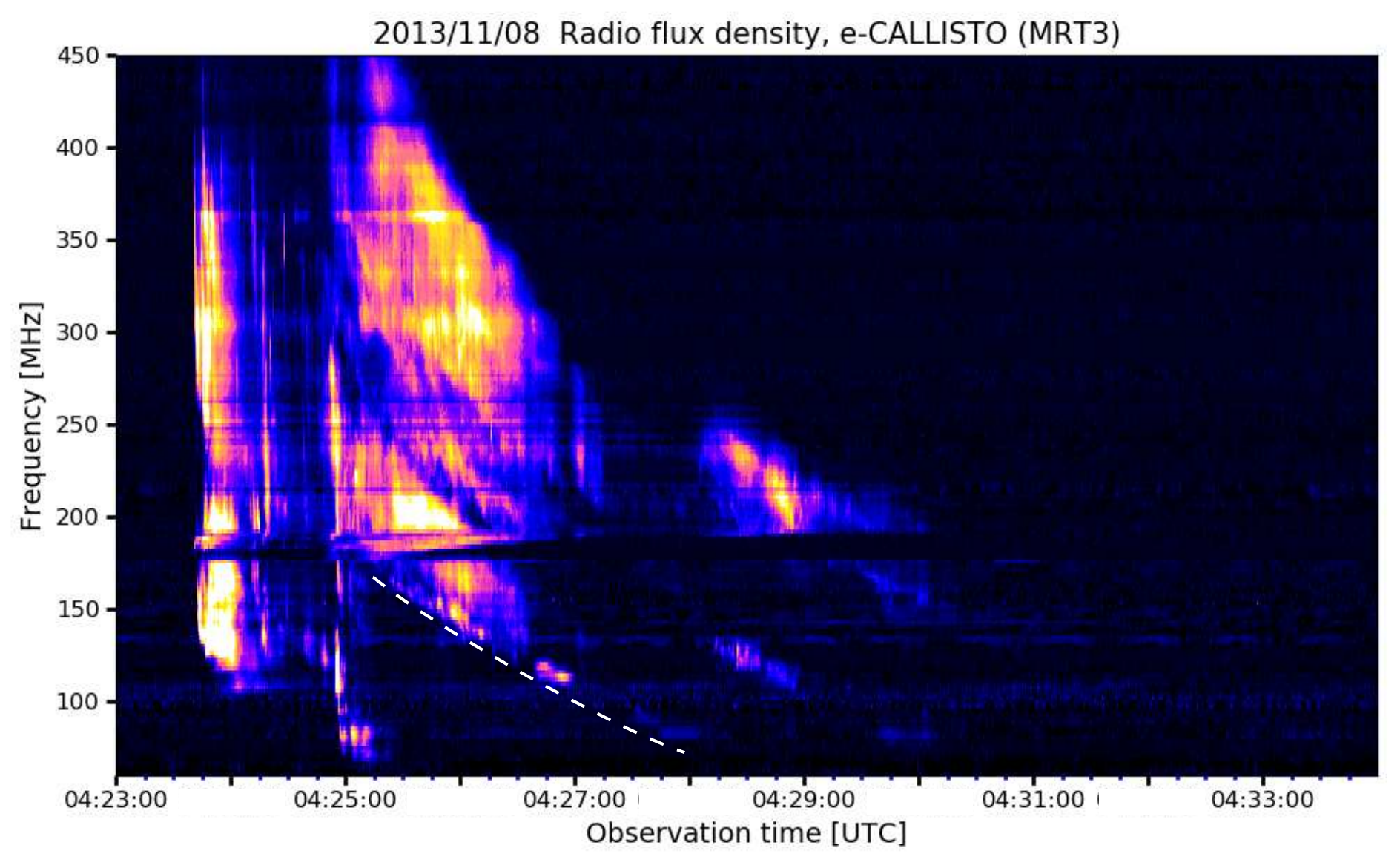}
\\ \hspace{5mm}  
\includegraphics[width=0.5\textwidth]{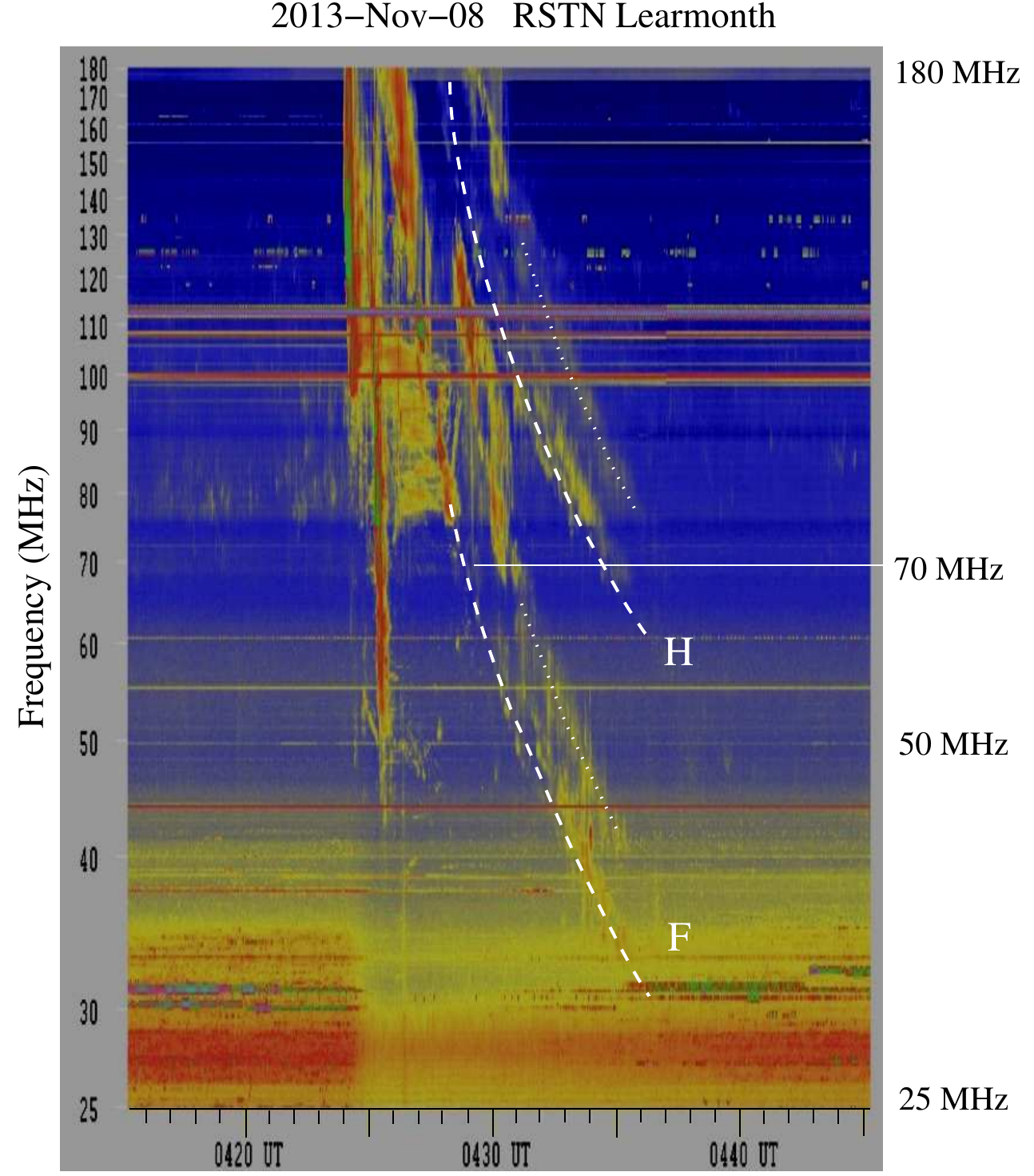}
\caption{Tracing the type II lanes in the e-CALLISTO MRT3 dynamic spectrum
  in the frequency range of 450--50~MHz (top) and in the RSTN Learmonth
  spectrum at 180--25~MHz (bottom). The fundamental (F) and harmonic (H)
  lanes are band-split (indicated with dashed and dotted lines in the 
  Learmonth spectrum, respectively).
  We used the fundamental, lower band-split lane in the height
  calculations. The times and heights are listed in Table \ref{table2}
  and they are shown in Figs.~\ref{fig:heights-low} and
  \ref{fig:radio-all}. The type II emission was preceded by type III bursts,
  which makes it difficult to define the exact type II burst start time.
 }
\label{fig:lear-times}
\end{figure}


\begin{figure}
\centering
\includegraphics[width=0.5\textwidth]{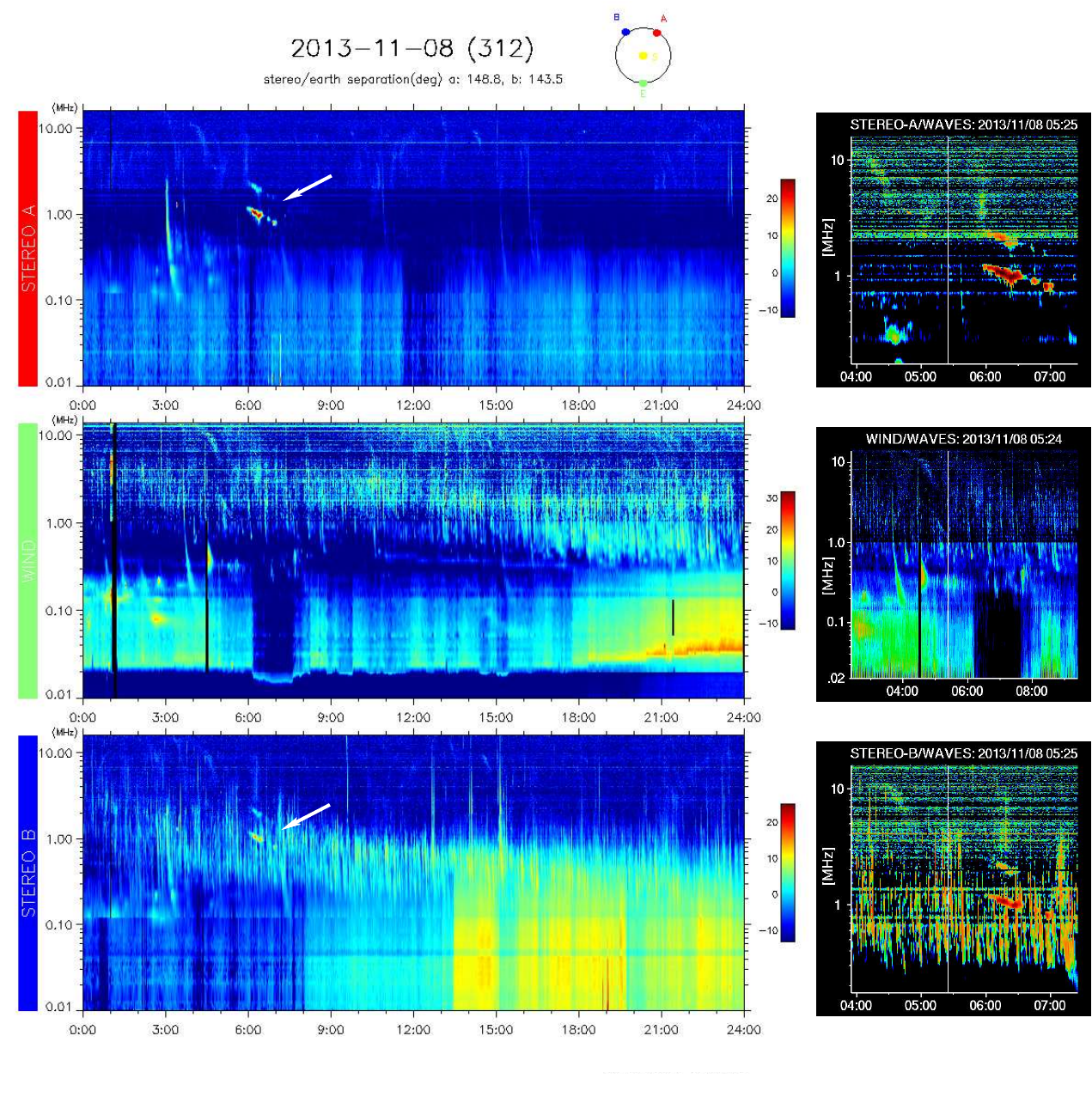}
\caption{STEREO A/SWAVES (location A in red), Wind/WAVES
  (location in green), and STEREO B/SWAVES (location B in blue)
  dynamic radio spectra from $\approx$\,10 MHz down to 10 kHz.
  The visible IP type II burst lanes are indicated with arrows. }
\label{fig:stereo}
\end{figure}


\subsection{Coronal mass ejections and particles}
\label{mass-ejection}

On the far side of the Sun both STEREO COR1 coronagraphs observed a
loop-shaped mass ejection at 03:05 UT (Fig. \ref{fig:backside-euvi}).
The eruption that started at 02:25 UT, near location S20W180, and the
following EUV dimmings match well with the CME shape and direction of
propagation.

On the earthside a CME was first observed by SOHO LASCO at 03:24 UT,
at height 2.7 R$_{\odot}$. The CME loop became fully visible in the next
available LASCO C2 image at 03:47~UT. The CME speed, determined from
LASCO observations, increased from 350 km s$^{-1}$ to 700 km s$^{-1}$
during the first seven hours of propagation (CDAW Catalog, 2nd order
fit to the CME leading front heights).

A separate plasmoid structure was observed inside the CME loop at 05:12 UT,
and it's front was fully visible in the difference image at 05:24 UT,
see Fig. \ref{fig:cme-flare} (the front is indicated with a green arrow).
The speed of this structure was estimated to be 400 km s$^{-1}$.
After 05:48 UT it could not be distinguished from the bright main loop
any longer.

\begin{figure}
\centering
\includegraphics[width=0.5\textwidth]{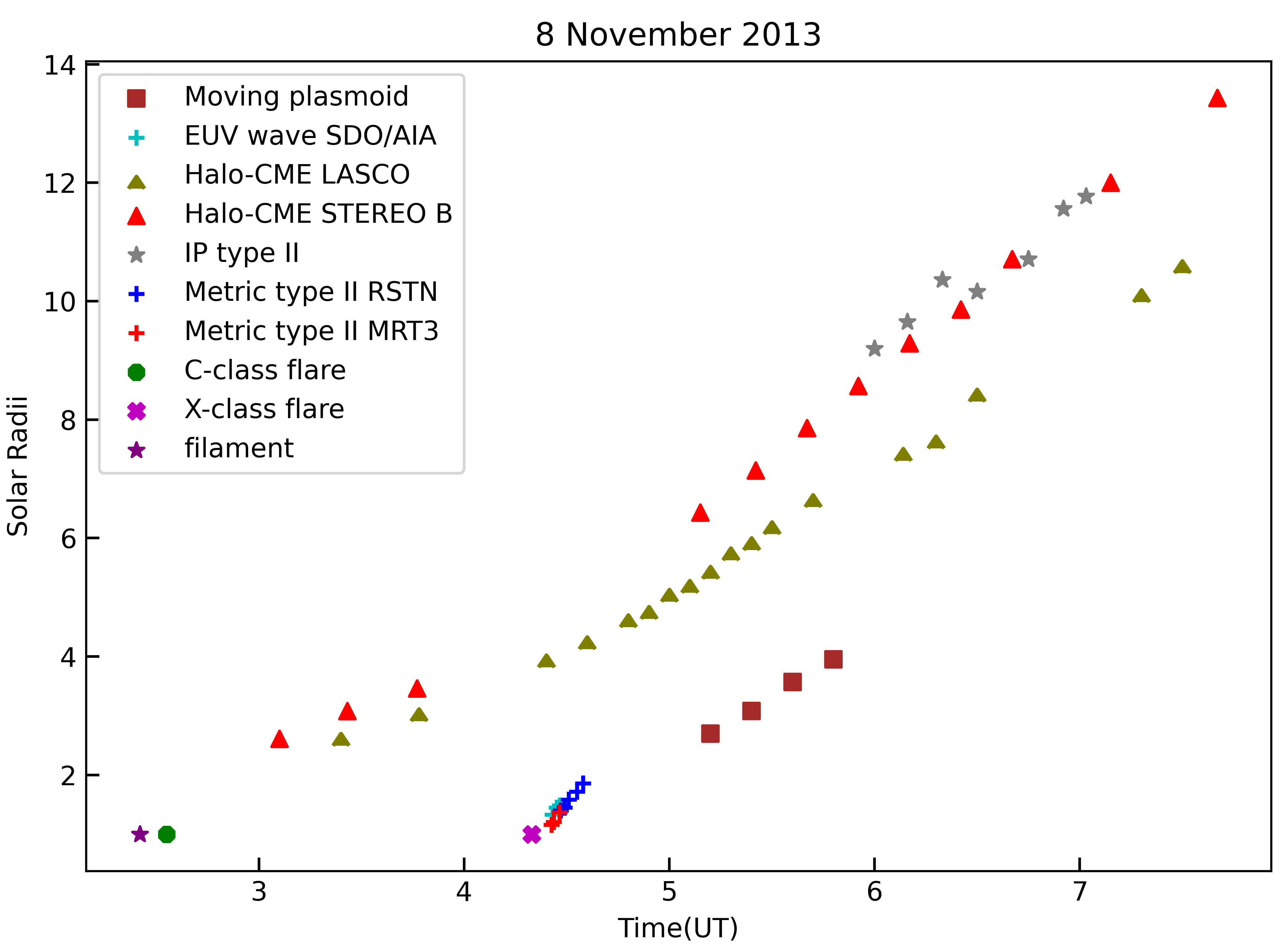}
\caption{Measured and calculated heights of all propagating structures.
  For flares and filament eruption we show their start times.
  }
\label{fig:radio-all}
\end{figure}

The CME and plasmoid heights are shown in Fig. \ref{fig:radio-all},
where they are presented together with the radio source heights that are
discussed in Section \ref{radio-emission}.

We checked in-situ particle measurements from both STEREO spacecraft
and Wind. A high-speed halo CME that was launched the day before,
near 10 UT on 7 November and originating also from the far side,
caused an increase in proton and electron intensities in STEREO A and B
particle detectors. The intensities remained high also on 8 November,
making it difficult to separate later enhancements. The 7 November
flare-CME-SEP event is described in detail in \cite{dresing2016}.



\begin{table}
\centering
\caption{Estimated type II burst heights. We used meter-wave observations
  from e-CALLISTO MRT3 and RSTN Learmonth, and DH-wave observations are from
  STEREO A and B SWAVES. Heights are heliocentric and they are calculated
  from emission at the fundamental plasma frequency, see text for details.
  }
\begin{tabular}{l|r|r}
Time     & Frequency & Height  \\
\hline
04:25:30 UT & 163 MHz  & 1.16 R$_{\odot}$  \\
04:25:56 UT & 147 MHz  & 1.18 R$_{\odot}$  \\
04:26:14 UT & 135 MHz  & 1.21 R$_{\odot}$  \\
04:26:55 UT & 111 MHz  & 1.26 R$_{\odot}$  \\
04:27:54 UT & 83  MHz & 1.37 R$_{\odot}$  \\
\hline
04:28:33 UT & 78.1 MHz & 1.40 R$_{\odot}$  \\
04:29:24 UT & 70.0 MHz & 1.45 R$_{\odot}$ \\
04:30:41 UT & 54.9 MHz & 1.58 R$_{\odot}$  \\
04:33:01 UT & 44.2 MHz & 1.72 R$_{\odot}$ \\
04:35:00 UT & 36.2 MHz & 1.86 R$_{\odot}$ \\
\hline
06:00 UT & 1.22 MHz & 9.20 R$_{\odot}$ \\
06:10 UT & 1.12 MHz & 9.65 R$_{\odot}$ \\
06:20 UT & 1.00 MHz & 10.36 R$_{\odot}$ \\
06:30 UT & 1.02 MHz & 10.16 R$_{\odot}$ \\
06:45 UT & 930 kHz & 10.71 R$_{\odot}$ \\
06:55 UT & 815 kHz & 11.56 R$_{\odot}$ \\
07:02 UT & 790 kHz & 11.77 R$_{\odot}$ 
\end{tabular}
\label{table2}
\end{table}

In the interplanetary space, below 15 MHz frequencies, very few type III
bursts are observed. Dynamic spectrum from STEREO A shows a type III burst
at 03:04 UT, first appearing at 2.5 MHz (height $\approx$\,6 R$_{\odot}$),
and Wind/WAVES shows a very faint type III burst appearing at 03:45 UT
at 4 MHz. This indicates that despite the flaring activity and CME
propagation, few accelerated particles had access to open field lines.
The Solar-MACH plot in Fig. \ref{fig:cme-flare} shows how the field
line from the X-class flare location is not connected to any of the
spacecraft, see \cite{gieseler2023} on how the visualization of the spatial
configuration and the solar magnetic connection to different observers
is done.
 
An IP type II burst was observed by STEREO A and B, appearing at 1.2 MHz
frequency at 06:00 UT. This type II burst was not observed by Wind
(located in L1 on the Earth-facing side), see Fig. \ref{fig:stereo}.
The type II burst emission lane could be followed down to 700 kHz,
where the emission ended around 07:00 UT. The calculated source heights,
using the fundamental emission lane and the same 'hybrid' density model,
are listed in Table \ref{table2}.

The radio source heights, with all other feature heights we have identified
and analysed, are plotted in Fig. \ref{fig:radio-all} for comparison. 
From this plot it is evident that the IP type II burst heights correspond
to the CME leading front heights observed from the far side, and that the
earthside LASCO observations of the CME fronts have lower heights.
Far side CME heights in Fig.~\ref{fig:radio-all} are from STEREO B,
as STEREO A observations show a wider CME structure, almost a double loop,
with several fronts. However, the heights of the leading parts are similar
to the STEREO B observations.
LASCO height measurements most probably suffer from projection effects,
as the halo CME originated from almost exactly the opposite side of the Sun.
This may well explain the lower heights.

The height-time evolution indicates that the metric type II burst and
the IP type II burst are not connected, but were created by separate
propagating shock waves.

\section{Summary and conclusions}
\label{summary}

We have analysed features and emissions that could have been associated
with a halo CME on 8 November 2013. We wanted to identify the origin
of the CME, and to exclude features that were not connected to it.
As eruptions were observed both on the earthside and on the far side of the
Sun, detailed analysis was needed. In halo CMEs the bright coronal emission
is observed in projection around the solar disk, and from coronagraph
images only it is impossible to conclude if the CME is launched toward
Earth or in the opposite direction. 

After investigating the eruptions and flares, some followed by EUV waves
and dimmings, and estimating the type II radio burst heights, we conclude
that the halo CME originated from the eruption on the far side of the Sun.
The IP type II burst was created by a shock wave (bow shock) ahead of the
halo CME, but the radio burst location was invisible to Earth.
The most probable reason for not observing it would be dense plasma
structures blocking wave propagation. An example of this kind of
configuration is presented in Fig. 11 in \cite{nasrin2018}.
If the radio source was located in a region directed towards STEREO A and B,
also the solar disc, or the CME itself, could have blocked radio emission
toward the Earth.

It is well known that type II bursts are typically formed in narrow
regions somewhere along the shock front, where plasma conditions are
favourable. The relatively short duration and narrow band of the
IP type II burst, and late appearance after the CME launch, suggest that
the radio emission could have been caused by interaction with a
small-volume ambient plasma structure.
 
During the CME propagation, the X-class flare eruption on the earthside of
the Sun caused a small plasmoid ejection, the material of which was
superposed on the halo CME features. The estimated heights of the metric
type II burst match well with the EUV wave launched by the X-class flare,
and the height-time trajectory of the plasmoid suggests they could be
associated. As the type II burst emission did not continue to IP space,
we conclude that either the shock died out (as a blast wave), or an increase
in the local Alfven speed made it impossible to propagate further out.


\section{Acknowledgments}
We thank the e-CALLISTO community for free access to their data and
software. MRT3 is located and operated at University of Mauritius,
Poste de Flacq, Mauritius.
The Radio Solar Telescope Network (RSTN) is a network of solar observatories
maintained and operated by the US Air Force. Learmonth is located in
Australia and their data is included in Space Weather databases.  
Nobeyama Radio Polarimeters (NoRP) are operated by Solar Science
Observatory, a branch of National Astronomical Observatory of Japan. 
The CME Catalog is generated and maintained at the CDAW Data Center by NASA
and the Catholic University of America in cooperation with the Naval Research
Laboratory. 


\end{document}